\documentclass[10pt,conference]{IEEEtran}
\usepackage{epsfig,setspace,amsmath,epsf,amssymb,bm,theorem,cite,graphicx, epstopdf,algorithm,algpseudocode,float,color,mathtools,authblk,tikz,physics}

\usetikzlibrary{positioning}
\usetikzlibrary{decorations.text}
\usetikzlibrary{decorations.pathmorphing}
\usepackage[short,c2]{optidef}

\usepackage{empheq}
\usepackage{cases}

\usepackage[table,xcdraw]{xcolor}
\usepackage{booktabs}
\usepackage{subfigure}
\usepackage{bbm}

\newtheorem{definition}{Definition}

\newtheorem{lemma}{Lemma}

\IEEEoverridecommandlockouts
\allowdisplaybreaks
\begin{document}
\title{Semi-Markov Decision Process Framework \\ 
for Age of Incorrect Information Minimization}

\author[1]{Ismail Cosandal}
\author[1]{Sennur Ulukus}
\author[2]{Nail Akar}

\affil[1]{\normalsize University of Maryland, College Park, MD, USA}
\affil[2]{\normalsize Bilkent University, Ankara, T\"{u}rkiye}

\maketitle
\begin{abstract}
For a remote estimation system, we study age of incorrect information (AoII), which is a recently proposed semantic-aware freshness metric. In particular, we assume an information source observing a discrete-time finite-state Markov chain (DTMC) and employing push-based transmissions of status update packets towards the monitor which is tasked with remote estimation of the source. The source-to-monitor channel delay is assumed to have a general discrete-time phase-type (DPH) distribution, whereas the zero-delay reverse channel ensures that the source has perfect information on AoII and the remote estimate. A multi-threshold transmission policy is employed where packet transmissions are initiated when the AoII process exceeds a threshold which may be different for each estimation value. In this general setting, our goal is to minimize the weighted sum of time average of an arbitrary function of AoII and estimation, and transmission costs, by suitable choice of the thresholds. We formulate the problem as a semi-Markov decision process (SMDP) with the {\em same state-space} as the original DTMC to obtain the optimum multi-threshold policy whereas the parameters of the SMDP are obtained by using a novel stochastic tool called \emph{dual-regime absorbing Markov chain} (DR-AMC), and its corresponding absorption time distribution named as \emph{dual-regime DPH} (DR-DPH). 
\end{abstract}

\section{Introduction}
Age of incorrect information (AoII) is a semantic-aware freshness metric proposed for remote estimation systems that measures the time the information at the monitor has been incorrect \cite{maatouk2020}. Different from the well-known age of information (AoI) metric \cite{Yates__HowOftenShouldone}, the AoII process can be brought down to zero without an update reception when the mismatch condition between the source and the monitor ends with a transition of the source to the estimated value at the monitor. Another difference from AoI is that, the AoII process can continue to rise even with an update reception if the source process has changed during the transmission of this update.

\begin{figure}[t]
	\centering
    \includegraphics[width=0.9\columnwidth]{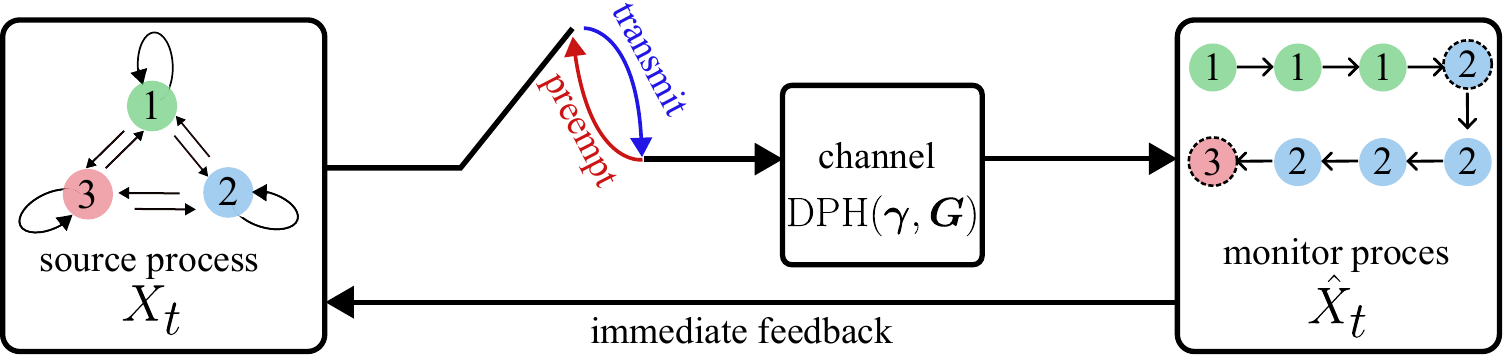}
	\caption{The remote estimation system involving the source process $X_t$, the monitor process $\hat{X}_t$, and the forward channel modeled by DPH($\bm{\gamma},\bm{G}$). The monitor updates its estimation with the received updates (marked with dashed circles).}
	\label{fig:SystemModel}
\end{figure}

We consider the remote estimation system in Fig.~\ref{fig:SystemModel} involving a source process $X_t$ and a monitor process $\hat{X}_t$ which is the remote estimate of $X_t$. The forward channel delay is modeled by a discrete phase type (DPH) distribution which is quite a general model that covers deterministic, geometric and mixed-geometric delays as sub-cases. The AoII process for this setting is defined by,
\begin{align}
      \text{AoII}_t & = t-\sup \{t^\prime: t' \leq t, X_{t'}=\hat{X}_{t'} \}, \label{eq:AoII}  
\end{align}
which considers a linear time penalty function with unit proportionality constant, a sub-case of the more general original formulation of AoII in \cite{maatouk2020}. Moreover, in this paper, we focus on minimizing the time average of an arbitrary function of both $\text{AoII}_t$ and the estimation value $\hat{X}_t$, denoted by $f_j(\text{AoII}_t)$ when $\hat{X}_t=j$, also named as the AoII penalty functions. A similar approach is used in \cite{xu2025timely}, where different age functions are used for nodes in a gossip network.

The AoII literature mainly focuses on special discrete-time Markov chain (DTMC) information sources \cite{maatouk2020, chen2021minimizing, maatouk2022age, 10778564}. These works propose a single-threshold policy which is obtained by formulating the optimization problem as a Markov decision process (MDP). To the best of our knowledge, \cite{cosandal2024modelingC} is the first work that considers a general continuous-time Markov chain (CTMC) source. In \cite{cosandal_etal_TRIT24}, it is proven for CTMC sources that the optimum transmission policy is a multi-threshold policy for which the threshold values depend both on the state of the source and the estimation value. However, such multi-threshold policies are shown in \cite{cosandal_etal_TRIT24} to be computationally very difficult to obtain for CTMCs with large state spaces. Therefore, in this paper, we propose to use multi-threshold policies, where the thresholds depend only on the remote estimate, and not on the source state. 

This paper considers an optimization problem where a weighted sum of AoII and transmission costs is minimized by employing an estimation-dependent multi-threshold policy for push-based transmissions. Particularly, an update packet is generated and transmitted when the AoII exceeds a threshold $\tau_j$ when the estimate is $\hat{X}_t=j$, and we seek the optimum thresholds $\tau_j$. For this purpose, the optimization problem is cast as a semi-Markov decision process (SMDP) \cite{ibe2013markov}, where the state-space of the SMDP amounts to the synchronization values at the embedded synchronization time points, with the time duration between these embedded points being random. For the purpose of obtaining the SMDP formulation, we adopt the multi-regime phase-type (MR-PH) distribution approach proposed for continuous-time problems in \cite{cosandal_etal_TRIT24} to be used for the discrete-time setting of the current paper. We note that phase-type (PH) distributions and absorbing Markov chains (AMC) have already been used in \cite{akar2023distribution, cosandal_etal_TRIT24, cosandal2024modelingC} to derive the distributions of AoI, peak AoI and AoII processes and they can be considered as an alternative to the stochastic hybrid systems (SHS) approach for AoI-related problems \cite{yates2020age}. Finally, once the SMDP formulation is complete, the optimum policy can be obtained by the policy iteration algorithm \cite{ibe2013markov}.

The contributions of this paper are summarized as follows. i)  AoII and sampling cost minimization problem is studied for an estimation-based penalty function under a general channel model. ii) The four-dimensional joint process involving the state of the original source, the estimation value, the channel phase, and the AoII process, is reduced to a one-dimensional embedded DTMC with the same state-space of the original source process. This enables us to find the optimum multi-threshold policy  efficiently. iii)  We propose the \emph{dual-regime AMC} (DR-AMC) and \emph{dual-regime discrete PH} (DR-DPH) analytical frameworks for two regimes that have a different number of transient states, to obtain the distribution of the AoII process, throughout the duration between two embedded synchronization time points.  

\subsection*{Notation}
Throughout the paper, we use bold lowercase and uppercase characters for a vector, and a matrix, respectively. Specifically, ${a}_{m}$ denotes the $m$th element of $\bm{a}$, and $a_{mn}$ denotes the $(m,n)$th element of $\bm{A}$. An $N\times N$ identity matrix is denoted by $\bm{I}_N$,  but the subscript can be omitted when the size is clear from the context. The vector $\bm{1}$ denotes a column vector of ones, and $\bm{e}_k$ denotes a column vector of zeros except for the $k$th entry, which is one. The operation `$\otimes$' corresponds to the Kronecker product \cite{linearAlgebra}.

\section{(Dual-regime) Phase-type Distributions and Dual-regime Absorbing Markov Chains}

\subsection{Discrete Phase Type Distribution}
A DPH-distributed random variable is defined as the time until absorption, denoted by $T$, in a finite-state discrete-time Markov chain (DTMC) $X_t, \ 1 \leq X_t \leq M+1, \ t\geq 0$. The first $M$ states (or phases) are the transient states, and the last state $M+1$ is the absorbing state \cite{telek_book}. The DTMC $X_t$ has the initial probability vector (IPV) of size $1 \times M$ denoted by $\bm{\beta}$, and the probability transition matrix 
$ \bm P= \Big( \begin{smallmatrix}
   \begin{array}{c|c} \bm{A} & \bm{a} \\ \hline  \bm{0} & 1 \end{array} 
\end{smallmatrix} \Big)$, 
for a sub-stochastic matrix $\bm{A}$ and a column vector $\bm{a}=\bm{1}-\bm{A}\bm{1}$, called the absorption vector.  In this case, we say $T \sim \text{DPH}(\bm{\beta},\bm{A})$ with order $M$. For more details on DPH distributions, see \cite{telek_book}.

\subsection{Dual-regime Absorbing Markov Chains}
We define an absorbing DTMC process $Y_t$ with $L \geq 1$ absorbing states, which is kicked off at time $t=0$. When the elapsed time is below the threshold $\tau$, i.e., $t < \tau$, then $Y_t$ is said to be in \emph{regime 1} with $K_1$ transient states. The IPV is denoted by the row vector $\bm{\beta}_1$ of size $K_1$. We denote the transient probability transition sub-matrix (TPTS) corresponding to the transition probabilities among the transient states by the square matrix $\bm{A}_1$ of size $K_1$. Also, we denote the absorbing probability transition sub-matrix (APTS), corresponding to the absorption probabilities from the transient states, by the $K_1 \times L$ matrix $\bm{B}_1$. If $Y_t$ is not absorbed for $t < \tau$, then \emph{regime 2} starts for which we have $K_2$ transient states. In addition, we define the \emph{boundary transition matrix} (BTM), denoted by $\bm{\Theta}$, a $K_1 \times K_2$ matrix, corresponding to the transition probabilities between transient states of regime 1 and those of regime 2, at the time instant $t = \tau-1$. Thus, the IPV of regime 2 can be written as,
\begin{align}
    \bm{\beta}_2=\bm{\beta}_1\bm{A}_1^{\tau-1}\bm{\Theta}.
\end{align}
Similar to regime 1, we denote the TPTS and APTS for regime 2 by the matrices $\bm{A}_2$ and $\bm{B}_2$ of sizes $K_2 \times K_2$ and $K_2 \times L$, respectively. Therefore, the probability transition matrix $\bm{P}_i$ governing the transitions of $Y_t$ is in the form $ \bm{P}_i= \Big( \begin{smallmatrix}
   \begin{array}{c|c} \bm{A}_i & \bm{B}_i\\ \hline  \bm{0} & \bm{I}_L \end{array} 
\end{smallmatrix} \Big), i=1,2$.

\subsection{Dual-regime Phase-type Distribution}
Consider the DR-AMC characterized with the 7-tuple $(\bm{\beta}_1,\tau,\bm{\Theta},\bm{A}_1,\bm{A}_1,\bm{B}_1,\bm{B}_2)$. 
The time to absorption of this DR-AMC is denoted by $T$ which is said to possess a DR-DPH distribution characterized with the 5-tuple $(\bm{\beta}_1,\tau,\bm{\Theta},\bm{A}_1,\bm{A}_2)$.

We will share a number of results regarding DR-AMCs and DR-DPH distributions without proof because of page limitations.
The probability mass function (pmf) of the random variable $T\sim \text{DR-DPH}(\bm{\beta}_1,\tau,\bm{\Theta},\bm{A}_1,\bm{A}_2)$ is given by,
\label{lem:drph}
\begin{align}
    p_T(t)& = \mathbb{P}(T=t) = \begin{cases}
    \bm{\beta}_{1}\bm{A}_{1}^{t-1}(\bm{1}-\bm{A}_{1}\bm{1}), & t<\tau, \\
    \bm{\beta}_{2}\bm{A}_{2}^{t-\tau}(\bm{1}-\bm{A}_{2}\bm{1}), & t\geq\tau. \label{eq:pt2}
    \end{cases}
\end{align}
We also define the absorption vector in regime $i$,  
$\bm{\sigma}_i  = \begin{pmatrix}
      \sigma_{i1} & \sigma_{i2} & \ldots & \sigma_{iL}  
\end{pmatrix}$,
where $\sigma_{i,j}$ is the probability of absorption into absorbing state $j$ out of a transition when the elapsed time resides in regime $i$. These two absorption vectors are given by
\begin{align}
    \bm{\sigma}_{1}&=\bm{\beta}_1 (\bm{I}-\bm{A}_1^{\tau-1})(\bm{I}-\bm{A}_1)^{-1}\bm{B}_1,  \label{eq:prob_dr1} \\
    \bm{\sigma}_{2}&=\bm{\beta}_2 (\bm{I}-\bm{A}_2)^{-1}\bm{B}_2.  \label{eq:prob_dr2}
\end{align}

\begin{lemma} \label{lem:pow}
    The $m$th factorial moment of $T$, namely $\nu_m=\mathbb{E}[T^{\underline{m}}], \ T^{\underline{m}}=T(T-1)(T-2) \cdots (T-m+1)$ \cite{graham1994concrete}, is given in closed form in \eqref{eq:pow}. Additionally, the ordinary moments of $T$, denoted by $\mu_T(m)=\mathbb{E}[T^{{m}}]$, can be obtained from the factorial moments \cite{bagui2024stirling} by,
    \begin{align}
        \mu_T(m)=\sum_{r=0}^m S(m,r) \nu_T(r), \label{eq:mom}
    \end{align}
    where $S(m,k)$ is the Stirling number of the second kind.
\end{lemma}

\begin{figure*}
        \begin{align}
        \nu_T(m)=\begin{cases}
            \bm{\beta}_1 \sum_{t=1}^{\tau-1} t^{\underline{m}}\bm{A}_1^{t-1}(\bm{1}-\bm{A}_1\bm{1})+\bm{\beta}_2 \sum_{r=0}^m \binom{m}{r} \tau^{\underline{m-r}} \ r!\bm{A}_2^r (\bm{I}-\bm{A}_2)^{-r-1}(\bm{1}-\bm{A}_2\bm{1}), & m\leq \tau, \\
            \bm{\beta}_1 \sum_{t=1}^{\tau-1} t^{\underline{m}}\bm{A}_1^{t-1}(\bm{1}-\bm{A}_1\bm{1})+ \bm{\beta}_2 \sum_{r=0}^m \binom{m}{r} m^{\underline{m-r}} \ r!\bm{A}_2^r (\bm{I}-\bm{A}_2)^{-r-1}   \bm{A}_2^{m-\tau}(\bm{1}-\bm{A}_2\bm{1}), & m> \tau, 
         \end{cases} \label{eq:pow}
    \end{align} 
    \hrulefill
\end{figure*}
    
We refer the reader to \cite{cosandal_etal_TRIT24} and \cite{akar_Asilomar} for further extensions of DR-AMC and DR-DPH to arbitrary number of regimes, in continuous and discrete time, respectively.

\section{System Model}
We consider a time-slotted remote estimation system with an $N$-state irreducible DTMC source process $X_t\in \mathcal{N}=\{1,2,\dots,N\}$ with the transition probability from state $i$ to state $j$ denoted by $q_{ij}$. In line with the generate-at-will (GAW) principle, the source can initiate a transmission of its observation to the remote monitor at the beginning of a time slot. We assume that state transitions of $X_t$ occur just before the end of the time slot, and when a transition occurs when there is an ongoing transmission, the source preempts the ongoing transmission, to avoid sending \emph{incorrect information}. We model the forward channel by $\text{DPH}(\bm{\gamma},\bm{G})$ with $M$ transient states with $\bm{h}=\bm{1}-\bm{G1}$. Remote monitor estimates the source process using the latest received information, i.e., $\hat{X}_t=X_{t'}$ where $t'$ is the generation time of the latest successful transmission. The mismatch between $X_t$ and $\hat{X}_t$ is measured with the AoII process defined in \eqref{eq:AoII}, where $f_j(\cdot)$ stands for the AoII penalty function for estimation $j$. We assume instantaneous feedback from the monitor to the source. Therefore, the source is always aware of the estimation and AoII processes. Consequently, we propose an estimation-based multi-threshold transmission policy for which a transmission is always initiated when the incorrect information stays longer than the threshold $\tau_j$ when the estimation $\hat{X}_t=j$. 

The objective of this work is to find a transmission policy for the source that minimizes the average cost,  
\begin{mini}
	{\substack{\bm{\tau}\in \mathbb{Z}_{+}^{N}}}
	{\lim_{L\to \infty}\frac{1}{L}\sum_{\ell=1}^L\text{cost}_\ell}
	{\label{Opt1}}
    {},
\end{mini}
where $\bm{\tau}$ is the vector of threshold values, and
\begin{align}
    \text{cost}_\ell =f_{\hat{X}_{\ell}}(\text{AoII}_\ell)+\lambda\delta_\ell, \label{eq:cost}
\end{align}
where $\delta_\ell$ is one if a transmission is ongoing at time $\ell$, and zero otherwise, and $\lambda$ denotes a relative weight assigned to the transmission cost component. 

\section{Semi-Markov Decision Process Formulation}
In order to obtain the SMDP formulation for the problem, we first need to define \emph{embedded points (EP).}

\begin{definition}[Embedded Points]
    A time point $t_0$ is called an embedded point (EP) with embedded value (EV) $E_j$ satisfying $\hat{X}(t_0) \neq  {X}(t_0)$, ${X}(t_0+1) = \hat{X}(t_0+1)=j$. Thus, embedded time points correspond to one time index before the source and monitor processes just get to synchronize at value $j$. 
\end{definition}

\begin{figure}[t]
    \centering
    \vspace{0.05in}
    \includegraphics[width=0.85\linewidth]{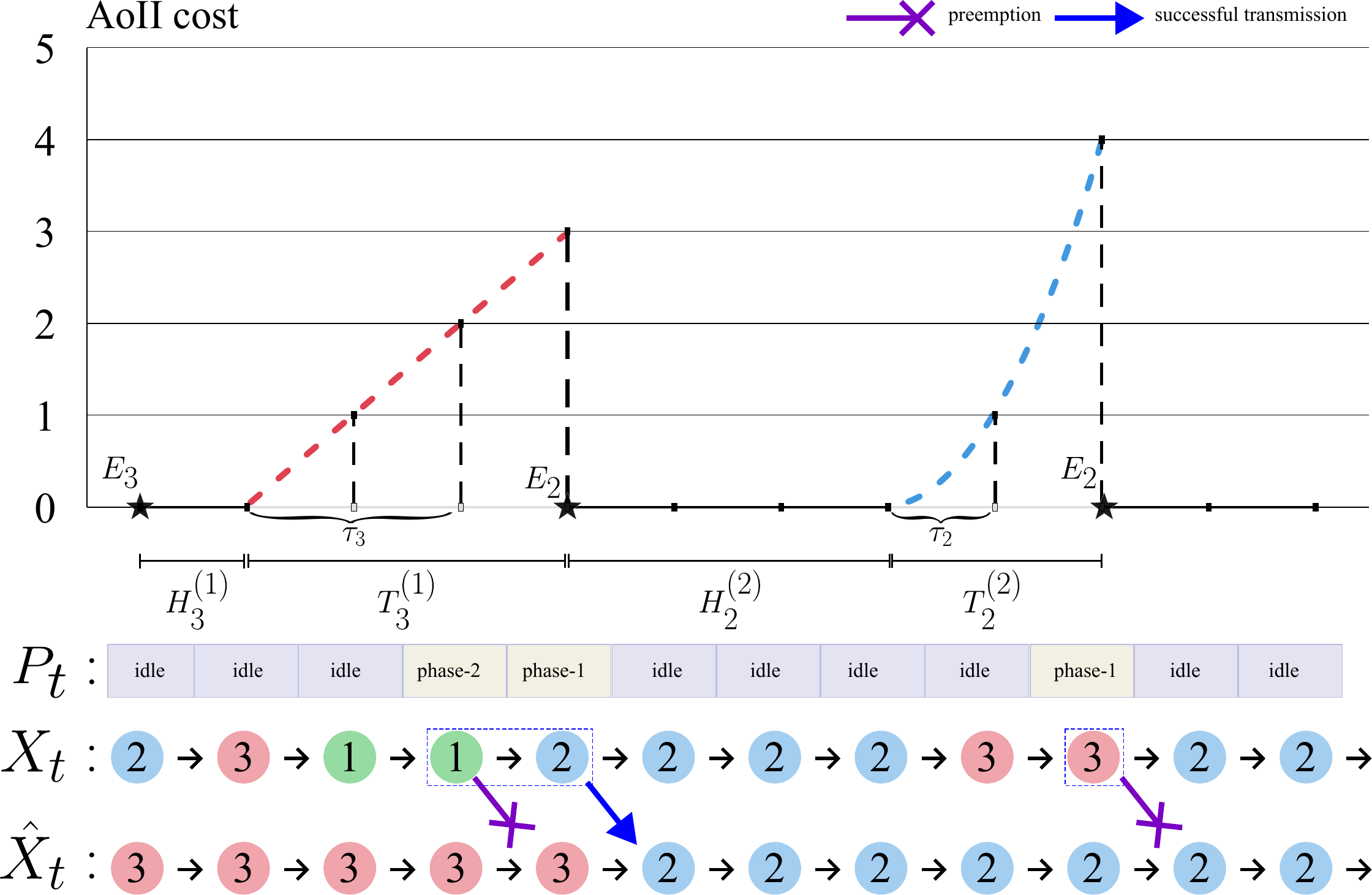}
    \caption{A sample path for the AoII cost throughout two complete cycles, when  $\tau_2=2$, $\tau_3=1$, and AoII penalty functions $f_2(x)=x^2$, $f_3(x)=x$ are used. $P_t$ amounts to the phase of the channel process in case of a transmission at time $t$. Otherwise, the channel is idle.}
    \label{fig:SP}
\end{figure}

The interval between the EP with EV $E_j$ and the next EP is called a cycle of type $j$, or cycle $j$ in short. Cycle $j$ is divided into two separate sub-intervals with the first one called the \emph{in-sync sub-interval} with its duration denoted $H_j$, which starts and lasts until the last time slot they stay in-sync. Then, the second sub-interval starts and lasts until the beginning of the next cycle. This sub-interval is called the \emph{out-of-sync interval} whose duration is denoted by $T_j$. During the out-of-sync interval, the source initiates a transmission whenever the duration exceeds the threshold $\tau_j$, which needs to be obtained for each $j$ to minimize the objective function in \eqref{Opt1}. Additionally, we denote the total duration of cycle $j$ by $D_j=H_j+T_j$. Fig.~\ref{fig:SP} depicts a sample path for the AoII cost throughout two cycles. 

Let us assume for the sake of simplicity that this cycle starts at $t=0$ at EV $j$, which gives rise to AoII cost $A_j=\sum_{t=1}^{T_j} f_j(t)$, and the total number of transmission attempts in cycle $j$ is denoted by $C_j$. Subsequently, the discrete-time semi-Markov decision process (SMDP) of interest to the current paper is the 5-tuple ($\mathcal{S},\mathbb{Z}_+,\rho,r,d$) \cite{ibe2013markov} described below.
\begin{itemize}
    \item Embedded values are the states of the problem with state space $\mathcal{S} = \{E_1,\ldots,E_N\}$.
    \item For each state $E_j$, the action is the value of the threshold $\tau_j \in \mathbb{Z}_+$, where $\mathbb{Z}_+$ denotes the set of positive integers, which then constitutes the action space of the problem.
    \item There are two costs of this problem which are, respectively, the \emph{age penalty} cost and the \emph{transmission} cost. For state $E_j$ and action $\tau_j$, we denote them with $a(E_j,\tau_j)=\mathbb{E}[A_j]$, and $c(E_j,\tau_j)=\mathbb{E}[C_j]$, respectively, and the expected total cost of the problem is denoted by $r(E_j,\tau_j)=a(E_j,\tau_j)+\lambda c(E_j,\tau_j)$. 
    \item Similarly, the expected duration for state $E_j$ for the same action equals $d(E_j,\tau_j)=\mathbb{E}[D_j]$.
    \item Lastly, $\rho(E_j,\tau_j,E_i)$ denotes the transition probability from state $E_j$ to next state $E_i$ when action $\tau_j$ is taken.  
\end{itemize}
A deterministic policy $\phi : S \rightarrow A$ is one that maps each state $s \in S$ to a single action $a \in A$, i.e., we take the action $\phi_s$ when we are at state $s$. The deterministic policy $\phi^*$ that minimizes the long-run average cost is the optimum solution for the SMDP problem, that can be obtained by the policy iteration algorithm \cite{book_tijms,ibe2013markov}. The SMDP parameters are calculated in the next section. 

\section{Calculation of the SMDP Parameters} \label{sec:calc}
In this section, we utilize the DR-AMC and DR-DPH frameworks which would allow us to calculate the parameters of the SMDP formulation. More specifically, we model the DTMC transitions from an EV $E_j$ to the next EV $E_i$ with these frameworks, and we then calculate the quantities $d(E_j,\tau_j)$, $a(E_j,\tau_j)$, and $\rho(E_i,\tau_j,E_j)$ for a given action $\tau_j$. Each cycle $j$ starts with an EV $E_j$ which indicates $X_t=\hat{X}_t$, and the two processes stay in-sync for a duration of $H_j$. During the in-sync interval, no transmission is initiated, and no penalty occurs, thus the total cost is zero. The synchronization is broken with the state change of $X_t$ to any state $i\neq j$, and the out-of-sync duration lasts until a new EP is reached. For each EV $j$, we define a DR-AMC process $Y_j(t)\sim\text{DR-AMC} (\bm{\beta}_{j1},\tau_j,\bm\Theta(\bm{\gamma}),\bm{A}_{j1},\bm{A}_{j2},\bm{B}_{j1},\bm{B}_{j2})$ to model the out-of-sync interval cycle $j$. Thus, the out-of-sync interval duration, denoted by $T_j$, has a DR-DPH distribution, i.e., $T_j\sim \text{DR-DPH}(\bm{\beta}_{j1},\tau_j,\bm\Theta(\bm{\gamma}),\bm{A}_{j1},\bm{A}_{j2})$.

Transient states of the first regime correspond to $i\in\mathcal{N}\backslash j$, enumerated as $\{1\dots,j-1,j+1,\dots,N\}$, and the EVs $E_i$, $i\in \mathcal{N}$ are the absorbing states. The first regime starts with the state transition of $j\to i$ of $X_t$ with the probability  $q_{ji}$, $j\neq i$. Thus, we can define the IPV for this regime with a row vector $\bm{\beta}_1$ whose elements can be calculated as $\{\bm{\beta}_{j1}\}_i= \frac{q_{ji}}{\sum_{k\neq j} q_{jk}}$, $i\neq j$. During this regime, the source does not initiate any transmission, thus only the absorbing state $E_j$ can be reached with a state transition of $X_t$ to the estimated value $j$. This regime lasts until an absorption to $E_j$ occurs, or $t$ reaches the threshold $\tau$. Table~\ref{tab:Tprob1} provides the transition probabilities from a transient state $i$ from which the matrices $\bm{A}_{j1}$ and $\bm{B}_{j1}$ can be constructed. 

The transient states of the second regime are denoted by $(i,m)$, where $i\in\mathcal{N}\backslash j$ and $m\in\{1,\dots,M\}$ correspond to the state of the source process and channel phase, respectively. We enumerate these states in the order $\{(1,1),(1,2),\dots,(j-1,M),(j+1,1),\dots (N,M)\}$. Absorbing states in the second regime are the same as in the first regime. When the elapsed time reaches the threshold $\tau_j$, then the source attempts a transmission visiting the channel phase $k$ with probability $\gamma_k$. Therefore, the BTM is obtained as $\bm{\Theta}(\bm{\gamma})=\bm{I}_{N-1} \otimes \bm{\gamma}$. Different from the first regime, the process can be absorbed to $E_i$, $i\neq j$ if i) the state $X_t$ does not change and ii) the transmission succeeds without preemption. These transition probabilities again are provided in Table~\ref{tab:Tprob2} from which the matrices $\bm{A}_{j2}$ and $\bm{B}_{j2}$ can be obtained similarly. 

\begin{table}[t]
    \caption{Transition probabilities for the process $Y_j(t)$ in the first regime.}
    \centering
    \begin{tabular}{|c|c|c|} 
   \hline
    \multicolumn{3}{|c|}{Transition probabilities from state $i$}\\ \hline
    To  & Condition & Probability\\ 
    \hline
    $i'$& $i'\neq i,j$ & $q_{ii'}$\\ \hline
    $E_j$& - &  $q_{ij}$  \\ \hline
    $E_{i}$&  $i'\neq j$ & $0$\\ \hline
    \end{tabular}
    \label{tab:Tprob1}
\end{table}

\begin{table}[t]
    \caption{Transition probabilities for the process $Y_j(t)$ in the second regime.}
    \centering
    \begin{tabular}{|c|c|c|} 
   \hline
    \multicolumn{3}{|c|}{Transition probabilities from state $(i,m)$}\\
    \hline
    To  & Condition & Probability\\ 
    \hline
    $(i,\ell)$& - & $q_{ii}g_{m\ell}$\\
    \hline
    $(i',\ell)$& $i'\neq i,j$ &$q_{ii'}\gamma_{\ell}$\\\hline
    $E_j$& - & $q_{ij}$  \\\hline
    $E_{i}$& $i'\neq j$& $(1-q_{ii})h_m$\\
    \hline
    \end{tabular}
    \label{tab:Tprob2}
\end{table}

\paragraph{Calculation of $a(E_j,\tau_j)$}
For any given AoII penalty function $f_j(t)$, the expected age cost can be calculated from the distribution in \eqref{eq:pt2} as
\begin{align}
    a(E_j,\tau_j) &= \mathbb{E}\left[\sum_{t=1}^{T_j} f_j(t) \right]=\sum_{t=1}^{T_j} f_j(t)p_{T_j}(t). \label{eq:aj}
\end{align}
Next, we provide the closed-form expression for $a(E_j,\tau_j)$ when the AoII penalty functions $f_j(t)$ are polynomial functions of $t$. 

\begin{lemma}[Polynomial AoII Penalty Functions]
    If the AoII penalty function for estimation value $j$ is polynomial with degree $K_j$, i.e., 
    $f_{j}(t)=\sum_{k=0}^{K_j} w_{k,j} t^k$ where $w_{k,j}$ are the polynomial coefficients, then the 
    following closed-form expression holds for $a(E_j,\tau_j)$,
    \begin{align}
         \sum_{k=0}^{K_j} w_{k,j}\sum_{m=1}^k \dfrac{S(m+1,n+1)}{n+1} \mu_{T_j}(m),\label{closedform}
    \end{align}
    where $\mu_{T_j}(m) = \mathbb{E}[T_j^m]$ can be obtained by \eqref{eq:mom}.
\end{lemma}
This expression can be obtained by Faulhaber's formula \cite{bagui2024stirling},
\begin{align}
    \mathbb{E}\left[\sum_{t=1}^{T}t^k\right]=\sum_{m=1}^k \dfrac{S(m+1,n+1)}{n+1} \mathbb{E}[T^m]. \label{eq:faul}
\end{align}
and using the relation between the ordinary and factorial moments in \eqref{eq:mom}.

\paragraph{Calculation of $d(E_j,\tau_j)$}
The expected duration of cycle $j$, denoted by $d(E_j,\tau_j)$ is the sum of $\mathbb{E}[H_j]$ and $\mathbb{E}[T_j]$. The former term equals the expected number of trials until success with the failure probability $q_{jj}$, which is $\mathbb{E}[H_j]=\frac{1}{q_{jj}}$. The second term can be calculated from \eqref{eq:pow} for $m=1$ by noticing $\nu_{T_j}(1)=\mu_{T_j}(1)$, which can be written as
\begin{align}
  & \sum_{t=1}^{\tau-1} t\bm{\beta}_{j1}\bm{A}_{j1}^{t-1}(\bm{1}-\bm{A}_{j1}\bm{1}) \nonumber\\ 
  &+\bm{\beta}_{j2}\left(\tau_j(\bm{I}-\bm{A}_{j2})^{-1}+\bm{A}_{j2}(\bm{I}-\bm{A}_{j2})^{-2} \right)(\bm{1}-\bm{A}_{j2}\bm{1}).  
\end{align}

\paragraph{Calculation of $c_{j}(\tau_j)$}
Since transmission happens each time slot in the second regime, the expected duration of the transmissions equals the number of state transitions in the second regime. From the fundamental matrix definition in \cite{cosandal_etal_TRIT24}, it can be calculated as $c(E_j,\tau_j)=\bm{\beta}_2(\bm{I}-\bm{A}_{j2})^{-1}\bm{1}$.

\paragraph{Calculation of $\rho(E_j,\tau_j,E_i)$}
The transition probability between embedded states $j\to i$ is equivalent to absorbing probabilities from absorbing state $E_j$ for the process $Y_{j}(t)$, and it can be calculated from \eqref{eq:prob_dr1} and \eqref{eq:prob_dr2} for each regime, separately. Alternatively, consider the absorbing states $j\neq i$, which only occur in the second regime. For the threshold value $\tau_j$, its absorbing probability can be calculated from \eqref{eq:prob_dr2} as
\begin{align}
   \rho(E_j,\tau_j,E_i)&=\bm{\beta}_{j2} (\bm{I}-\bm{A}_{j2})^{-1}\bm{B}_{j2}\bm{e}_i, \quad i\neq j.
\end{align}
Then, the self-transition probability for the $E_j$ can be found by $\rho(E_j,\tau_j,E_j)=1-\sum_{k\neq j}\rho(E_j,\tau_j,E_k)$.

\section{Numerical Results}
In this section, we present numerical examples for validating the SMDP model of the paper along with comparisons with two benchmark policies: i) \emph{Single-threshold (ST) policy} for which the source waits for a single system-wide threshold $\tau$ while there is a mismatch between $X_t$ and $\hat{X}_t$, which is a widely used mechanism in the literature \cite{chen2021minimizing, maatouk2020}, and the value of $\tau$ which minimizes the overall cost is found by line search, ii) \emph{Random sampling (RS) policy} \cite{cosandal_etal_TRIT24} in which the source initiates a transmission with probability $\xi$ when there is a mismatch, and again the value of $\xi$ which minimizes the overall cost for RS is found by line search.

We study the following scenarios. In scenario I, the information source has $N=10$ states, and has a probability transition matrix $\bm{Q}$ whose diagonal elements are linearly spread in the interval $[0.4,0.6]$, and whose off-diagonal elements are linearly spread
in the interval $[0.5\frac{1-q_{nn}}{N-1}, 1.5\frac{1-q_{nn}}{N-1}]$. For this source, we consider the AoII penalty functions $f_j(x)=\frac{1}{j}x^2+\frac{1}{N+1-j}x$ for estimation $j$. For this scenario, we consider a geometrically distributed channel delay with parameter  $0.8$, equivalently $\bm{G}=0.2$, $\gamma=1$.
In scenario II, we verify our analytical approach for
for a 3-state original process $\bm{Q}=\begin{bsmallmatrix}
        0.60 & 0.25 & 0.15 \\
        0.25 & 0.55 & 0.20 \\
        0.20 & 0.30 & 0.50
        \end{bsmallmatrix}$
with the channel delay distributed according to $\text{DPH}(\gamma,\bm{G})$ with $\bm{\gamma}=\begin{bmatrix}
            1 & 0 \\ 
        \end{bmatrix}$, $\bm{G}=\begin{bsmallmatrix}
            0.7 & 0.2 \\ 0.1 & 0.6
        \end{bsmallmatrix}$, 
 and the linear AoII penalty functions $f_1(x)=x+\frac{1}{2}$, $f_2(x)=\frac{1}{2}x+1$, $f_3(x)=\frac{1}{3}x+\frac{1}{4}$ are used.

\begin{figure}
    \begin{center}
    \subfigure[]{\includegraphics[width=0.45\linewidth]{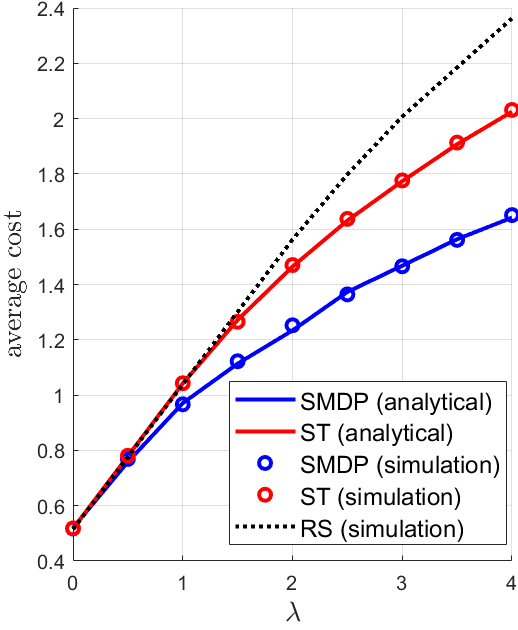}} ~ 
    \subfigure[]{\includegraphics[width=0.45\linewidth]{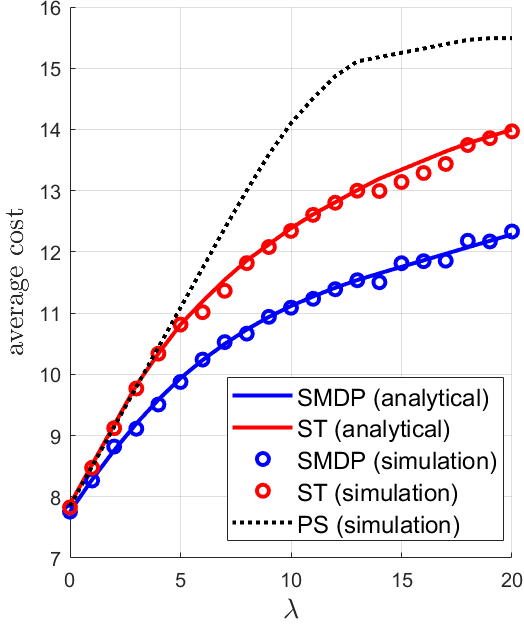}} 
    \end{center}  
    \caption{Comparison of benchmark policies with proposed SMDP policy with varying $\lambda$ for a) scenario I b) scenario II.}
    \label{fig:gen}
\end{figure}

For a given weight parameter $\lambda$, the proposed SMDP algorithm obtains the multi-threshold optimum policy. The average cost obtained with the SMDP algorithm along with the RS and ST policies is depicted in Fig.~\ref{fig:gen} for both scenarios.
For ST and SMDP policies, both analytical and simulation results are obtained, and the agreement between them verifies our analytical results. Since all AoII penalty functions are polynomials in the examples, we employed the closed-form expression \eqref{closedform} for finding the parameters of the SMDP model. On the other hand, only simulation results are used for RS. We observe that our proposed SMDP-based policy outperforms both benchmark policies substantially, and all converge to the same policy when $\lambda \rightarrow 0$ which corresponds to the {\em always transmit} policy.

\bibliographystyle{IEEEtran}
\bibliography{bibl}

\end{document}